\documentclass{article}
\usepackage{float}
\usepackage{authblk}

\title{A New Initial Distribution for Quantum Generative Adversarial Networks to Load Probability Distributions}

\author[1]{Yuichi Sano\thanks{\texttt{sano.yuichi.77v@st.kyoto-u.ac.jp}}} 
\affil[1]{Department of Nuclear Engineering, Kyoto University, Nishikyo-ku, Kyoto 615-8540, Japan} 
\author[1]{Ryosuke Koga} 
\author[2]{Masaya Abe} 
\affil[2]{Nomura Asset Management Co.,Ltd.} 
\author[2]{Kei Nakagawa} 

\date{\today}

\usepackage{booktabs}
\usepackage{multirow}

\usepackage{graphicx}
\usepackage{amsmath,amsthm,amssymb}
\usepackage{amsfonts}
\usepackage{siunitx}

\usepackage{subcaption}

\usepackage{algorithm, algpseudocode}

\usepackage{qcircuit}
\usepackage{braket}
\usepackage{cite}
\usepackage{comment}

\theoremstyle{definition}

\algtext*{EndWhile}
\algtext*{EndIf}

\algrenewcommand\algorithmicrequire{\textbf{Input:}}
\algrenewcommand\algorithmicensure{\textbf{Output:}}

\begin{document}

\maketitle

\begin{abstract}
Quantum computers are gaining attention for their ability to solve certain problems faster than classical computers, and one example is the quantum expectation estimation algorithm that accelerates the widely-used Monte Carlo method in fields such as finance.
A previous study has shown that quantum generative adversarial networks(qGANs), a quantum circuit version of generative adversarial networks(GANs), can generate the probability distribution necessary for the quantum expectation estimation algorithm in shallow quantum circuits. 
However, a previous study has also suggested that the convergence speed and accuracy of the generated distribution can vary greatly depending on the initial distribution of qGANs' generator.%the generator of qGANs.
In particular, the effectiveness of using a normal distribution as the initial distribution has been claimed, but it requires a deep quantum circuit, which may lose the advantage of qGANs. 

Therefore, in this study, we propose a novel method for generating an initial distribution that improves the learning efficiency of qGANs.%the efficiency of qGANs learning. 
Our method uses the classical process of {\it label replacement} to generate various probability distributions in shallow quantum circuits. 
We demonstrate that our proposed method can generate %the log-normal distribution, which is important in financial engineering, the triangular distribution, and the bimodal distribution more efficiently than existing methods.
the log-normal distribution, which is pivotal in financial engineering, as well as the triangular distribution and the bimodal distribution, more efficiently than current methods.

Additionally, we show that the initial distribution proposed in our research is related to the problem of determining the initial weights for qGANs.
\end{abstract}

\section{Introduction}
%Backgroud
Quantum computers have attracted attention for their usefulness in solving specific problems faster than classical computers, and in recent years they have been the subject of very active research and development in various fields\cite{Shor,Grover}.
%One area where the use of quantum computers could be particularly useful is in fields that make extensive use of Monte Carlo methods, such as financial engineering.
One area where quantum computing could be particularly beneficial is in fields that rely heavily on Monte Carlo methods, such as financial engineering. 
%Since quantum computers have an algorithm for fast expectation estimation (quantum expectation estimation algorithm), 
Since quantum computers can utilize a fast expectation estimation algorithm, known as the quantum expectation estimation algorithm, it is expected that the algorithm can be used instead of the Monte Carlo method in classical computers to speed up expectation estimation\cite{review_quantum_finance,Rebentrost18,Egger20}.

However, it is known that %Noisy Intermediate-Scale Quantum(NISQ) devices currently being developed 
current Noisy Intermediate-Scale Quantum (NISQ) devices under development are noisy due to the lack of error correction. %and that 
Additionally, there are limits to the circuit depth, which corresponds to how many gates can be executed in a quantum circuit, and the number of qubits corresponding to a size of a computation\cite{arute2019quantum,NISQ_review}.
Therefore, there are many ongoing studies to improve algorithms to make existing quantum algorithms viable for NISQ devices\cite{peruzzo2014variational,zoufal2019quantum}.

%Previous Research
%In order to perform the quantum expectation estimation algorithm, it is necessary to construct a quantum circuit that loads a probability distribution into a quantum state.
Executing the quantum expectation estimation algorithm needs the construction of a quantum circuit capable of loading a probability distribution into a quantum state. 
One method is the Grover-Rudolph state preparation\cite{grover2002creating}, but it requires %a maximum of
up to $O(2^n)$ gates and quantum circuit depth, making it difficult to use in NISQ devices.
To address this issue, a method using quantum Generative Adversarial Networks(qGANs), which are quantum circuit versions of Generative Adversarial Networks(GANs\cite{goodfellow14}), was proposed in a previous research\cite{zoufal2019quantum}.
Previous research has shown that qGANs can be used to learn the parameters of quantum circuits to generate shallow quantum circuits that load probability distributions into quantum states.
On the other hand, the same previous research suggested that the speed of learning and the accuracy of the generated distributions differed greatly depending on the initial distribution used for a qGANs' generator.
In particular, the effectiveness of using the normal distribution as the initial distribution was argued, but using the normal distribution requires more depth of the quantum circuit.
In the previous research, the circuit with a depth of ``$d=4$'' could make the normal distribution, including errors, the initial distribution in the case of $3$ qubits.
However, as the number of qubits increases, a deeper quantum circuit is considered necessary to generate a normal distribution.
Therefore, qGANs using the normal distribution for the initial distribution may lose the advantage of shallow circuit depth, a feature of circuit generation using qGANs.

%Our study and result
We propose a method for creating initial distributions that can be implemented in shallow quantum circuit depth, enhancing the learning efficiency of qGANs.%making the learning of qGANs more efficient.
Specifically, we propose a method to generate various probability distributions even for circuits with gate depth $d=1$, similar to the uniform distribution, by a classical process of {\it label replacement}.
Since the only operations required to execute the proposed method can be performed on a classical computer, it can be efficiently implemented on a quantum computer, including NISQ devices.%for example, even on NISQ devices.
We then propose a method to create an initial distribution ``near'' the target distribution using only single-qubit gates by {\it label replacement}.
To confirm the effectiveness of our method, %we verified by simulation that 
we conducted simulations to verify that the log-normal, triangular, and bimodal distributions investigated in previous studies can be generated more efficiently and accurately than previous initial distributions.
We have shown by simulation that our method performs better than the uniform distribution consisting of single-qubit gates for all target distributions and better than the normal distribution for the log-normal distribution, which is essential in financial engineering.
We also showed that the initial distribution proposed in this study is the initial weights of the generator and, consequently, can be implemented at depth $d=0$.

\section{Preliminaries}
In this section, we outline the method of loading probability distributions into quantum states using qGANs, which is the basis of this study.%, and its application which is the quantum expectation estimation algorithm.
Furthermore, we detail its practical application within the context of the quantum expectation estimation algorithm.

\subsection{Quantum Expectation Estimation Algorithm}
Monte Carlo-based expectation estimation using a classical computer requires $\mathcal{O}(1/\epsilon^2)$ number of simulations for the desired accuracy $\epsilon$.
In contrast, the quantum expectation estimation algorithm, which uses an amplitude estimation algorithm, is known to be able to estimate an expectation in $\mathcal{O}(1/\epsilon)$ oracle queries\cite{Rebentrost18}.

The quantum expectation estimation algorithm consists of two steps: (1) preparing a quantum state whose probability amplitude has an expected value of an estimation target and (2) estimating its amplitude.
We show how to load the target expectation into a probability amplitude of a quantum state.
First, we construct the following oracle that loads the target probability into the probability amplitude;
\begin{equation}
    \mathcal{P} \ket{0}^{\otimes n} = \sum_{i}^N \sqrt{p_i}\ket{i},
\end{equation}
where $i$ is each event, $p_i$ is the probability of event $i$ occurring, and $n$ is the number of qubits such that the number of all events $N=2^n$.
For this quantum state, the probability of measuring $\ket{1}$ when the second register is measured in a computational basis is
\begin{equation}
\text{Pr}[\ket{1}]=\sum_{i}^N f_i p_i
\end{equation}
which is the expected value of the target.
Therefore, an oracle $\mathcal{Q}=\mathcal{P F}$ that combines these two oracles $\mathcal{P}$ and $\mathcal{F}$ is an oracle that loads the desired expected value into a probability amplitude of a quantum state.

We then show quantum amplitude estimation to estimate the probability amplitude using this oracle $\mathcal{Q}$ and the quantum phase estimation algorithm\cite{nielsen2002quantum}.
Recently, algorithms have been proposed to avoid the inverse quantum Fourier transform, which has a deep circuit\cite{Suzuki2020,Grinko2021}.
We simplify the notation of the quantum state after running the oracle $\mathcal{Q}$ as follows;
\begin{equation}
    \mathcal{Q}\ket{0}^{\otimes(n+1)} = \cos{\theta} \ket{\psi_0}\ket{0} + \sin{\theta}\ket{\psi_1}\ket{1}.
\end{equation}
The probability amplitude we want to find is $\sin{\theta}$.
In addition to the oracle $\mathcal{Q}$, oracles $\mathcal{S}_0=I-2\ket{0}\bra{0}^{\otimes(n+1)}$, and $\mathcal{S}_{\psi_0}=I-2\ket{\psi_0}\bra{\psi_0}\otimes\ket{0}\bra{0}$ are combined to form a oracle $\mathcal{U} = \mathcal{Q}\mathcal{S}_0\mathcal{Q}^{\dagger}\mathcal{S}_{\psi_0}$.
Since the phase of the oracle $\mathcal{U}$ is $\pm{e^{i2\theta}}$, the quantum phase estimation algorithm can estimate $\theta$ and $\sin{\theta}$ with an error $\epsilon$ by querying $\mathcal{O}(1/\epsilon)$ times for the oracle $\mathcal{Q}$.
As mentioned above, this $\sin{\theta}$ is the desired expected value in the quantum expectation estimation algorithm, so the computational complexity is improved compared to using the Monte Carlo method.

\subsection{Loading Probability Distributions by qGANs}

\begin{figure}[tb]
    \centering
    \includegraphics[width=\linewidth]{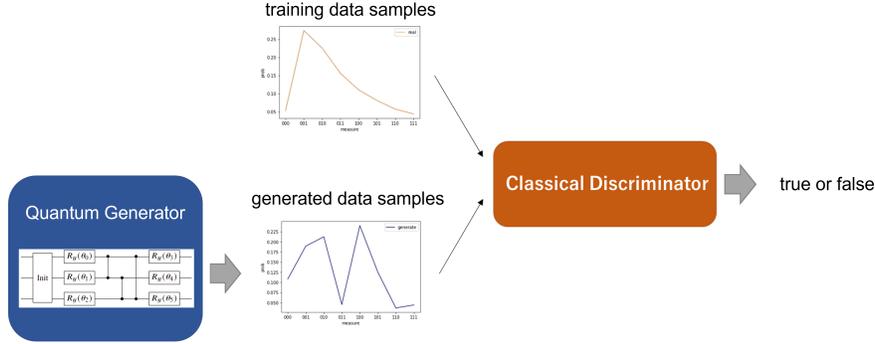}
    \caption{Loading Probability Distributions by qGANs}
    \label{fig:qGAN}
\end{figure}
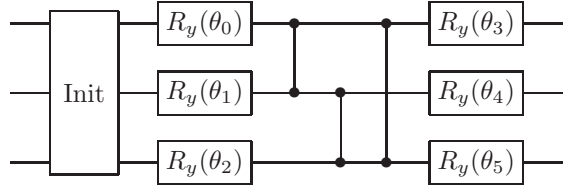
\begin{figure}[tb]
\begin{equation*}
     \Qcircuit @C=1.5em @R=1em {
  & & & & \multigate{2}{\rm{Init}}  & \gate{R_y(\theta_0)}	 & \ctrl{1}  & \qw &  \ctrl{2} &\gate{R_y(\theta_3)} &  \qw \\
   &  & & &  \ghost{\rm{Init}}   & \gate{R_y(\theta_1)}	 & \ctrl{0}  & \ctrl{1}& \qw & \gate{R_y(\theta_4)}  &  \qw\\
    & & & &  \ghost{\rm{Init}}   & \gate{R_y(\theta_2)}	 & \qw  & \ctrl{0} & \ctrl{0}  & \gate{R_y(\theta_5)} &  \qw
    }
\end{equation*}
\caption{A quantum circuit with parameters that serve as a quantum generator. The circuit $\rm{Init}$ is the circuit that generates an initial distribution for learning.}
\label{fig:qcircuit}
\end{figure}

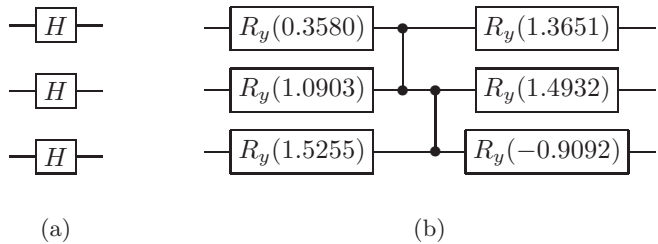
\begin{figure}[tb]
  \begin{minipage}[b]{0.3\linewidth}
    \centering
    \begin{equation*}
    \Qcircuit @C=1em @R=1.2em {
    &  \gate{H}	&   \qw \\
    &   \gate{H}	 &  \qw\\
    &    \gate{H}	  &  \qw
    }
    \end{equation*}
    \subcaption{}    
    \label{fig:init_u}
  \end{minipage}
  \begin{minipage}[b]{0.5\linewidth}
    \centering
    \begin{equation*}
    \Qcircuit @C=1em @R=.7em {
    & \gate{R_y(0.3580)}& \ctrl{1} & \qw &  \gate{R_y(1.3651)}	&   \qw \\
    & \gate{R_y(1.0903)} & \ctrl{0} &\ctrl{1}  &   \gate{R_y(1.4932)}	 &  \qw\\
    & \gate{R_y(1.5255)} & \qw  & \ctrl{0} &    \gate{R_y(-0.9092)}	  &  \qw
    }
    \end{equation*}
    \subcaption{}    
    \label{fig:init_n}
  \end{minipage}
  \caption{Circuits to generate initial distributions. (a) Circuit to generate the uniform distribution. (b) Circuit to generate the normal distribution with mean $\mu=1$ and variance $\sigma=1$( for log-normal distribution).}
\end{figure}

In order to run the quantum expectation estimation algorithm, it is necessary to construct the oracle $\mathcal{P}$ that loads a probability distribution into a probability amplitude.
It is known that using the method Grover-Rudolph state preparation\cite{grover2002creating} for this purpose requires the use of up to $\mathcal{O}(2^n)$ gates, which leads to a deep circuit depth.
Therefore, to construct oracles that generate probability distributions with shallower quantum circuits, the method using qGANs was proposed in previous research\cite{zoufal2019quantum}.

Generative Adversarial Networks (GANs) are generative models used to create false data nearly indistinguishable from the true data they are based.
In GANs, the generator, which generates false data, and the discriminator, which discriminates between false and true data, learn from each other so that, eventually, an adequately trained generator will output false data indistinguishable from true data.
The qGANs are an extension to a generator by replacing the generator with a quantum circuit, as shown in Figure \ref{fig:qGAN}.
In previous research, a method was proposed to create a quantum circuit that prepares a quantum state with probability amplitude indistinguishable from the target probability distribution by providing data sampled from the target probability distribution as true data for the qGANs.
The quantum generator is a quantum circuit with parameters such that the rotation angle $\theta_i$ of the $R_y$ gate is varied as a ``weight'', as shown in Figure \ref{fig:qcircuit}.
The $R_y(\theta_i)$ gates represent the rotation of the y-axis of the Bloch sphere and correspond to the following unitary operator;
\begin{equation}
R_y(\theta_i) = 
\begin{pmatrix}
\cos{\frac{\theta_i}{2}} & -\sin{\frac{\theta_i}{2}} \\
\sin{\frac{\theta_i}{2}} & \cos{\frac{\theta_i}{2}} \\
\end{pmatrix}.
\end{equation}

\section{Method}
\begin{figure}[tb]
  \begin{minipage}[b]{0.7\linewidth}
    \centering
    \includegraphics[width=70mm]{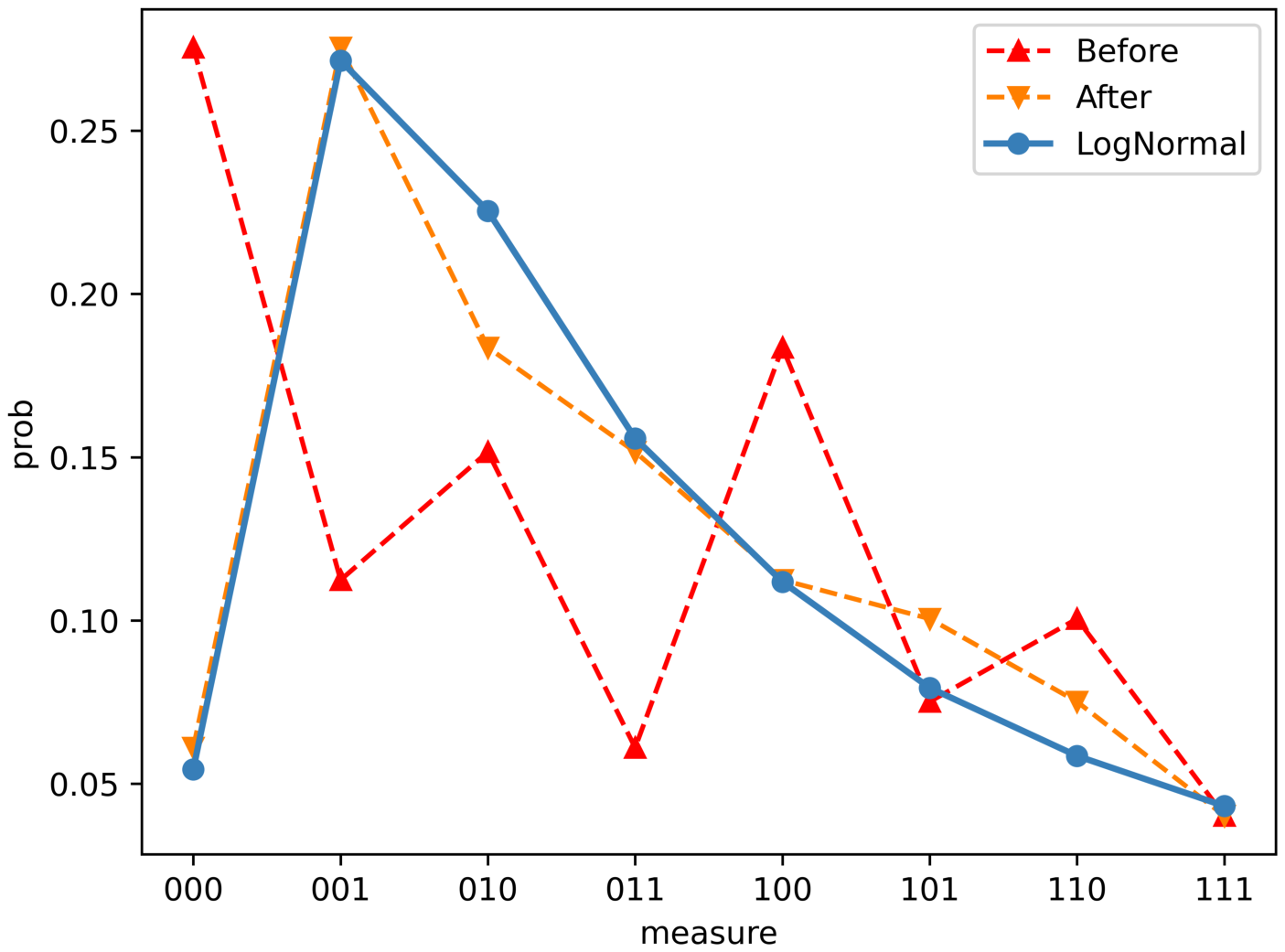}
    \subcaption{}
    \label{fig:label}
  \end{minipage}
  \begin{minipage}[b]{0.2\linewidth}
    \centering
    \begin{equation*}
    \Qcircuit @C=1em @R=2.4em {
    &  \gate{R_y(\theta_0 = 0.36 \pi)}	&   \qw \\
    &   \gate{R_y(\theta_1 = 0.40 \pi)}	 &  \qw\\
    &    \gate{R_y(\theta_2 = 0.44 \pi)}	  &  \qw
    }
    \end{equation*}
    \subcaption{}    
    \label{fig:init}
  \end{minipage}
  \caption{Transformation of the initial distribution by {\it label replacement}. (a) ``Before'' is the distribution before {\it label replacement}, ``After'' is the distribution after {\it label replacement}, and ``LogNormal'' is the distribution that approximates the target log-normal distribution(mean $\mu=1$ and variance $\sigma=1$). (b) The specific quantum gate of the quantum circuit $\rm{Init}$ loads the distribution before {\it label replacement}.}
\end{figure}

Our purpose is to %propose 
introduce a method for creating initial distributions that uses only single-qubit gates and is more efficient in learning qGANs.

For this purpose, we propose {\it label replacement} that can create various initial distributions from a probability distribution that can be generated from single-qubit gates.
Specifically, {\it label replacement} reproduces various initial distributions without increasing the number of quantum gates by appropriately replacing the measurement results on a computational basis.
For example, by replacing Figure \ref{fig:label}, we reproduce an initial distribution close to the target log-normal distribution.
The circuit $\rm{Init}$ for generating the distribution before the replacement of Figure \ref{fig:label} can be implemented with a gate of depth $d=1$ composed by Figure \ref{fig:init}.
If one were to consider replacing the measurement results with only operations on the quantum circuit, the replacement would require two-qubit gates, and it is impossible to implement at a depth of $d=1$.
In contrast, {\it label replacement} can be implemented by a classical process because it transforms a random variable that determines what value corresponds to the output of a measurement result.
Therefore, this method is easy to implement in any quantum circuit, including NISQ devices.

We then show how to create an initial distribution that reflects knowledge using {\it label replacement}.
To prepare an initial distribution that is ``near'' to the target probability distribution, we first sort the target probability distribution in order of increasing probability.
Next, we determine the angle of $R_y$ so that the sum of the probabilities of the top half corresponds to the measured probability of $\ket{0}$.
Then we determine the angle of $R_y$ so that the ratio of the sum of its further top half probabilities corresponds to the measurement probability of $\ket{0}$.
We repeat the above procedure to determine the angle of $R_y$.
Using this procedure, we can always generate an initial distribution such that the largest probability in the target probability distribution coincides with the largest probability in the probability distribution created by {\it label replacement}.
We present the exact algorithm in Algorithm \ref{alg:al1}.

\begin{algorithm}[t]
\caption{Fitting Probability Distributions}
\label{alg:al1}
\begin{algorithmic}
    \Require $\{p_n\}_n$
    \Comment{Input is a target probability distribution.}
    \Ensure $\theta_{target}$
    \Comment{Output is the rotation angle of the $R_y$ gates.}
    \State $\{q_n\}_n \gets \text{Sort}(\{p_n\}_n) (\therefore q_i \geq q_{i-1})$
    \Comment{Sort $p_n$.}
    \State $u_1 \gets 1$
    \State $\theta_{target} = []$
    \For{$i=1 \text{ \bf to }\log(n)$}
    \Comment{Calculate the rotation angle $\theta_i$.}
    \State $u_i = \Sigma_{j=1}^{n/2^i} q_j$
    \State $r_i = u_i/u_{i-1}$
    \State $\theta_i = 2 \arccos(r_i) \quad (0 \leq \theta_i \leq \frac{\pi}{2})$
    \State $\theta_{target}.\text{append}(\theta_i)$
    \Comment{Add the obtained rotation angle to the list $\theta_{target}$.}
    \EndFor 
\end{algorithmic}
\end{algorithm}

\section{Simulation Test}
The performance of our initial distributions was tested by simulating a quantum computer with a classical computer.

\subsection{Simulation Setup}
Our simulation setup was the same as previous research \cite{zoufal2019quantum} except that we added our initial distribution.
We also used log-normal, triangular, and bimodal distributions to evaluate performance.
The log-normal distribution had a mean $\mu = 1$ and a standard deviation $\sigma = 1$, the triangular distribution had a lower limit $l = 0$, an upper limit $u = 7$, and mode $m = 2$, and the bimodal distribution consisted of two superimposed Gaussian distributions with $\mu_1 = 0.5$, $\sigma_1 = 1$ and $\mu_2 = 3.5$, $\sigma_2 = 0.5$.
The quantum generator of the qGANs consists of an initial distribution circuit $init$ and gates with parameters as shown in Figure \ref{fig:qGAN}.
The circuit with parameters for the latter part of the quantum generator consists of a repeating quantum circuit with one set of combinations of a section of $R_y$ gates with the rotation angle $\theta_i$ as weights and a section of $CZ$ gates to affect multiple qubits.
The iterations of these gates correspond to the number of parameters, with $N_p = 3(k+1)$ number of parameters for $k$ number of iterations.
The initial weights of the generators were randomly sampled values from a uniform distribution in the range $[-10^{-1}, 10^{-1}]$.
The classical discriminator was implemented using Pytorch\cite{paszke2019pytorch}.
The discriminator consists of a 50-node input layer, a 20-node hidden layer, and a 1-node output layer.
The AMSGRAD\cite{ReddiKK18} was used for training, with a learning rate of $lr=10^{-4}$ and coefficients for calculating moments of a gradient of $\beta_1=0.7$ and $\beta_2=0.99$.
The parameter shift rule \cite{Mitarai18,Zeng19} was used to calculate the gradient of the quantum generator, and the number of samples was set to 8000 times.
The following non-saturating loss function\cite{Fedus18} is used for a loss function:
\begin{equation}
\mathcal{L}_{G}(\phi, \theta) = - \mathbb{E}_{z \sim p_{\text{prior}}} \left[ \log D_\phi (G_\theta(z)) \right],
\end{equation}
\begin{eqnarray}
\mathcal{L}_D(\phi, \theta) &=& \mathbb{E}_{x \sim p_{\text{real}}} \left[ \log D_\phi(x) \right] + \mathbb{E}_{z \sim p_{\text{prior}}} \left[ \log(1 - D_\phi(G_\theta(z))) \right],
\end{eqnarray}
where $p_{prior}$ denote the true probability distribution of $z$,  $p_{real}$ denote the probability distribution obtained from the data, $\mathcal{L}_D$ is the loss function of the discriminator and $\mathcal{L}_G$ is the loss function of the generator.

As training data, we used samples from each target probability distribution, sampled 2000 times, and discretized by 8 points.
These data were divided into mini-batches of batch size 2000, and with the addition of 2000 samples measured from the quantum generators, we trained the qGANs.
The training was performed up to 2000 epochs, with ten training runs each for each initial distribution and each target probability distribution to ensure robustness.

In this study, the initial distribution was our initial distribution, the uniform distribution, and the normal distribution with a mean and standard deviation of the target probability distribution.
Let $\theta_{target} = [\theta_0, \theta_1, \theta_2]$ be the parameters that make up our initial distribution, as shown in Figure \ref{fig:init}.
The parameters of our initial distribution are 
\begin{equation}
    \theta_{lognormal} = [0.36\pi, 0.40\pi, 0.44\pi]
\end{equation}
for the log-normal distribution,
\begin{equation}
    \theta_{triangular} = [0.29\pi, 0.43\pi, 0.48\pi]
\end{equation}
for the triangular distribution, and
\begin{equation}
    \theta_{bimodal} = [0.21\pi, 0.44\pi, 0.48\pi]
\end{equation}
for the bimodal distribution.
Also, the permutation of each measurement outcome is
\begin{equation}
    \sigma_{lognormal} = 
\begin{bmatrix}
0 & 1 & 2 & 3 & 4 & 5 & 6 & 7 \\
3 & 0 & 4 & 2 & 1 & 6 & 5 & 7 \\
\end{bmatrix}
\end{equation}
for the log-normal distribution,
\begin{equation}
    \sigma_{triangular} =
\begin{bmatrix}
0 & 1 & 2 & 3 & 4 & 5 & 6 & 7 \\
3 & 6 & 0 & 4 & 2 & 1 & 5 & 7 \\
\end{bmatrix}
\end{equation}
for the triangular distribution, and
\begin{equation}
    \sigma_{bimodal} =
\begin{bmatrix}
0 & 1 & 2 & 3 & 4 & 5 & 6 & 7 \\
2 & 6 & 1 & 0 & 4 & 5 & 3 & 7 \\
\end{bmatrix}
\end{equation}
for the bimodal distribution.
We tested for the case $k={1,2,3}$ as the number of iterations of the quantum circuit.

For the simulation of a quantum computer, we used qiskit\cite{qiskit}, a quantum computation library provided by IBM.

We use relative entropy to measures to evaluate the training performance.
Relative entropy is a type of statistical distance that measures the closeness of the discrete probability distribution $P(x)$ and $Q(x)$ and is given by
\begin{align}
    H(P|Q):= \sum_{x \in X} P(x) \log (\frac{P(x)}{Q(x)}).
\end{align}
Relative entropy is non-negative, and the probability distributions $P(x)$ and $Q(x)$ are equal if and only if the value is zero.

\subsection{Result}
\begin{table}[htbp]
\centering
\caption{Relative entropy of the result of learning the target distribution for ten times: The table shows the initial distributions(Init), the repeat($k$), the mean ($\mu$), the standard deviation ($\sigma$), and the minimum(Min).}
\begin{tabular}{cccccc}
\hline
 Data & Init & $k$ & $\mu$ & $\sigma$ & Min \\ \hline
 \multirow{9}{*}{log-normal} &\multirow{3}{*}{our method}  & 1 & 0.00190 & 0.000290 & 0.00134  \\
 & & 2 & 0.00163 & 0.000405 & 0.00092  \\
  &          & 3 & 0.00163 & 0.000514 & 0.00086  \\ \cline{2-6}
 & \multirow{3}{*}{uniform} & 1 & 0.14514 & 0.013234 & 0.12992 \\
 &  & 2 & 0.07810 & 0.043166 & 0.03962 \\ 
 & & 3 & 0.07702 & 0.052596 & 0.00706 \\ \cline{2-6}
  & \multirow{3}{*}{normal}& 1 & 0.00414 & 0.005282 & 0.00129  \\
 & & 2 & 0.01260 & 0.009894 & 0.00119  \\
 & & 3 & 0.01320 & 0.018007 & 0.00097  \\\hline
 
 \multirow{9}{*}{triangular} &\multirow{3}{*}{our method}  & 1 & 0.03962 & 0.007225 & 0.03078  \\
 & & 2 & 0.03931 & 0.011143 & 0.00744  \\
  &          & 3 & 0.02480 & 0.017297 & 0.00165  \\ \cline{2-6}
 & \multirow{3}{*}{uniform} & 1 & 0.36796 & 0.129145 & 0.06295 \\
 &  & 2 & 0.08249 & 0.068867 & 0.01054 \\ 
 & & 3 & 0.19945 & 0.264557 & 0.00274 \\ \cline{2-6}
 & \multirow{3}{*}{normal}& 1 & 0.00212 & 0.000903 & 0.00136  \\
 & & 2 & 0.00289 & 0.003174 & 0.00104  \\
 & & 3 & 0.00288 & 0.001490 & 0.00121  \\\hline
 
 \multirow{9}{*}{bimodal} &\multirow{3}{*}{our method}  & 1 & 0.17156 & 0.041284 & 0.10309  \\
 & & 2 & 0.11749 & 0.077465 & 0.00474  \\
  &          & 3 & 0.09782 & 0.067075 & 0.00194  \\ \cline{2-6}
 & \multirow{3}{*}{uniform} & 1 & 0.34285 & 0.015677 & 0.31533 \\
 &  & 2 & 0.16386 & 0.060814 & 0.05680 \\ 
 & & 3 & 0.04484 & 0.039178 & 0.01275 \\ \cline{2-6}
 & \multirow{3}{*}{normal}& 1 & 0.00953 & 0.001066 & 0.00786 \\
 & & 2 & 0.00846 & 0.002848 & 0.00100  \\
 & & 3 & 0.00863 & 0.003180 & 0.00191  \\\hline
\end{tabular}
\label{tab:result}
\end{table}

The results of the simulation are summarized in Table \ref{tab:result}.
From Table \ref{tab:result}, our method is superior to the uniform distribution as initial distribution in both mean and minimum values of relative entropy, except for the mean value of $k = 3$ for the bimodal distribution.
Our method and the uniform distribution consist of single-qubit gates, which means that our method is a better initial distribution for the same cost.
Also, our method is inferior in the mean value of relative entropy but equally suitable for the minimum value compared to the case where the normal distribution is an initial distribution.
In particular, when the log-normal distribution is the target distribution, our method outperforms the normal distribution as the initial distribution in both mean and minimum values, indicating the superiority of our method.
Our method may be better in NISQ devices because the normal distribution requires two-qubit gates, which is expensive due to the high noise content in NISQ devices, and the depth is deeper than our method.
Our method is also robust to the qubit number scale because it is always ``$d = 1$'', whereas normal distributions may lose the advantage of circuit generation using qGANs if the quantum circuit making the normal distribution is deeper for more than four qubits.

\begin{figure}[tb]
\begin{minipage}[b]{0.32\columnwidth}
\centering
\includegraphics[width=\linewidth]{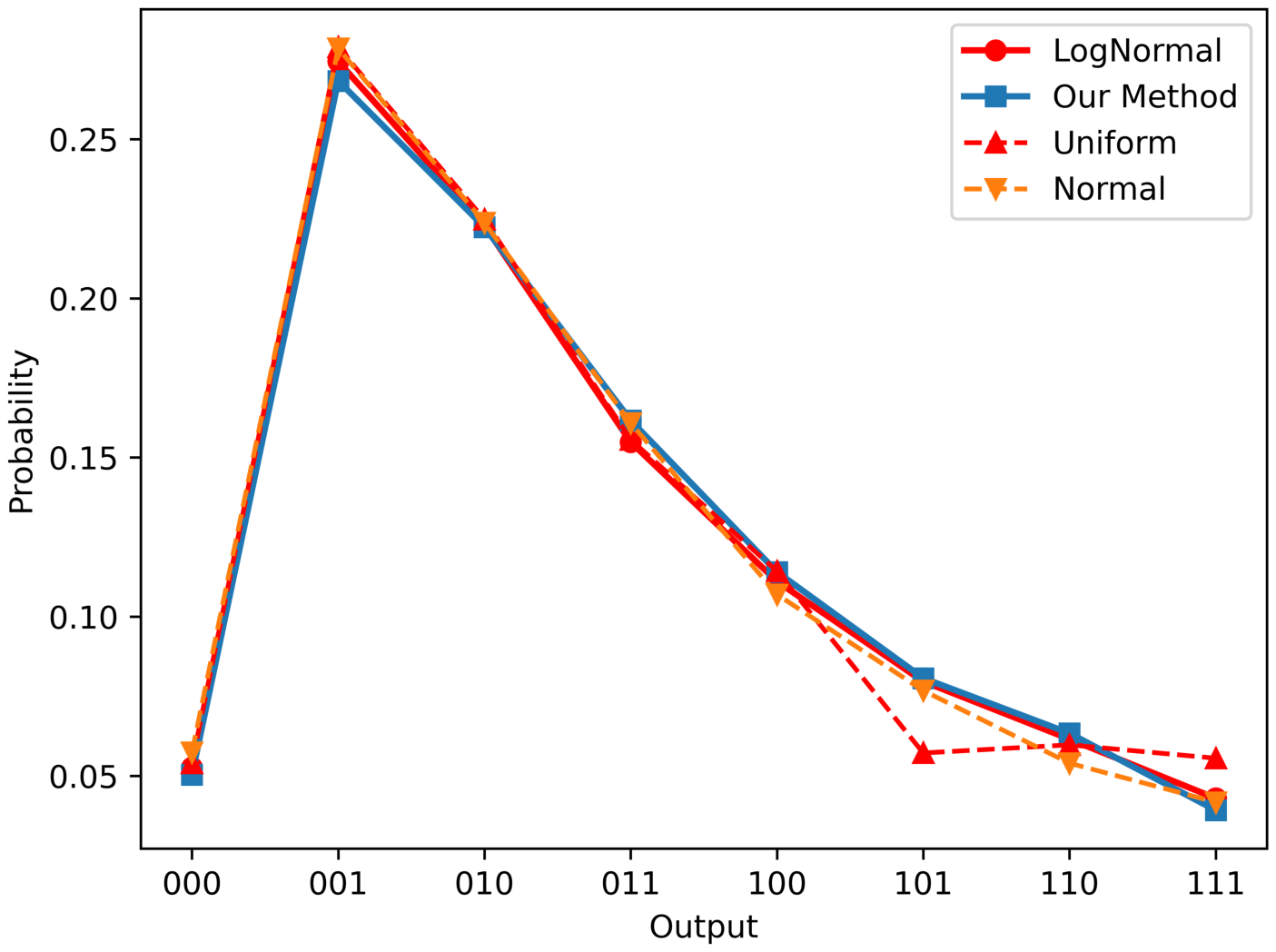}
\subcaption{log-normal}
\end{minipage}
\begin{minipage}[b]{0.32\columnwidth}
\centering
\includegraphics[width=\linewidth]{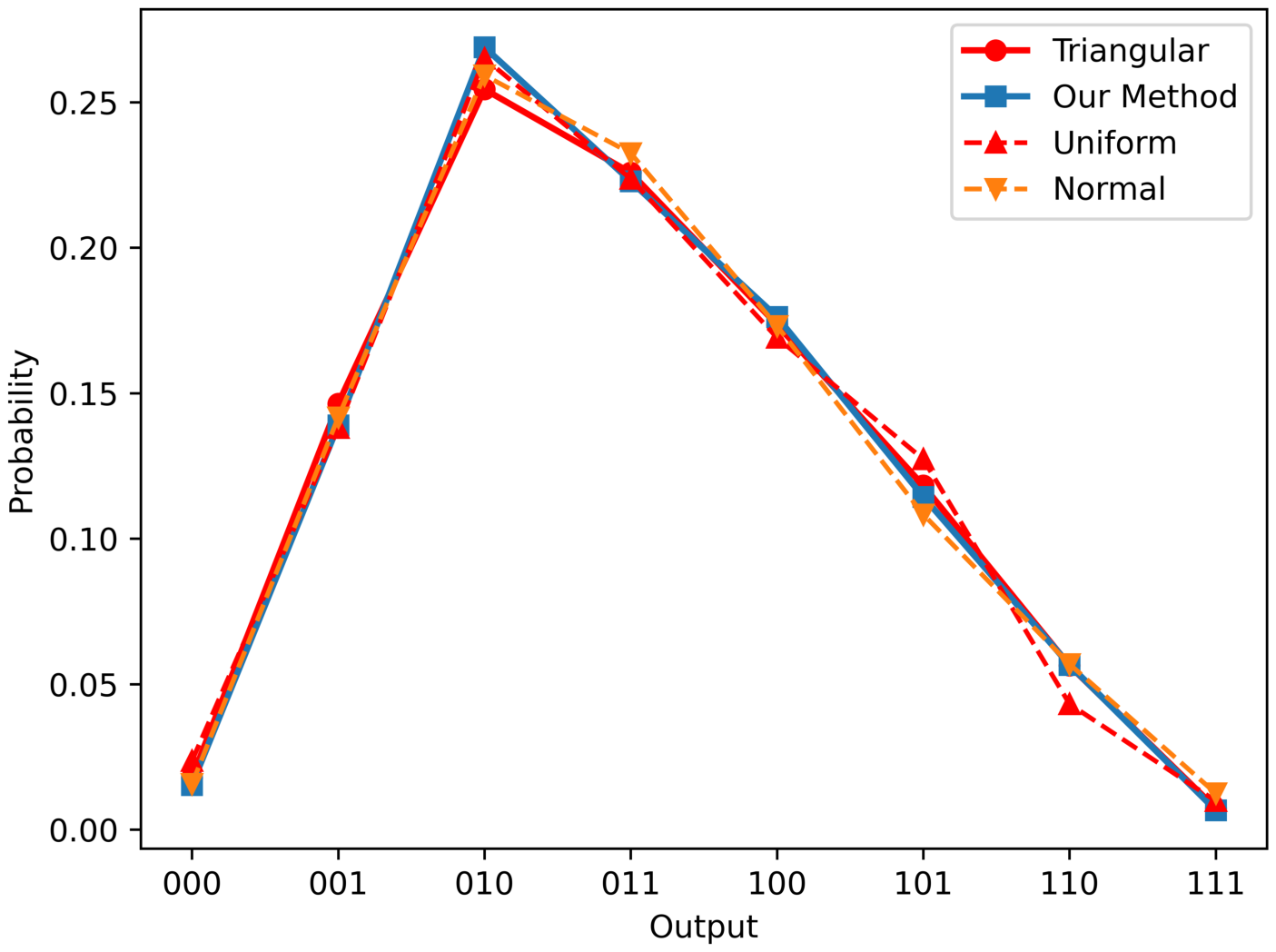}
\subcaption{triangular}
\end{minipage}
\begin{minipage}[b]{0.32\columnwidth}
\centering
\includegraphics[width=\linewidth]{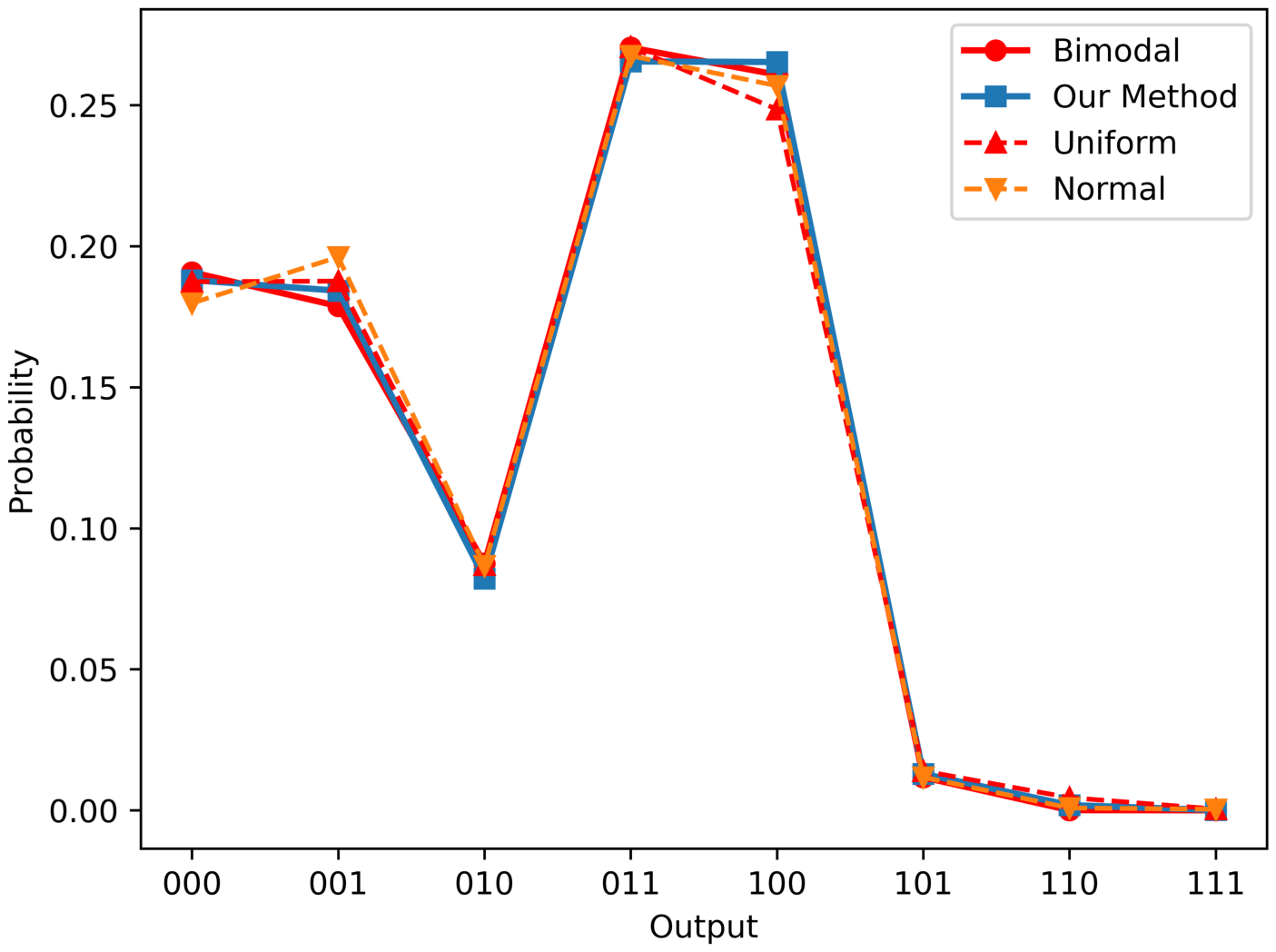}
\subcaption{bimodal}
\end{minipage}
\caption{Comparison of generated probability distributions. The lowest relative entropy for our proposed method (``Our Method''), the uniform distribution (``Uniform''), and the normal distribution (``Normal'') with $k = 3$.}
\label{fig:distribution}
\end{figure}

As a legend, the probability distribution with the lowest relative entropy generated from each initial distribution is shown in Figure \ref{fig:distribution}.

\subsection{Relationship with Initial Weight}

\begin{figure}[tb]
\begin{minipage}[b]{\linewidth}
\centering
\begin{equation*}
     \Qcircuit @C=1.5em @R=1em {
& \gate{R_y(0.36 \pi)}  & \gate{R_y(\theta_0)}	 & \ctrl{1}  & \qw &  \ctrl{2} &\gate{R_y(\theta_3)} &  \qw \\
& \gate{R_y(0.40 \pi)}   & \gate{R_y(\theta_1)}	 & \ctrl{0}  & \ctrl{1}& \qw & \gate{R_y(\theta_4)}  &  \qw\\
& \gate{R_y(0.44 \pi)}  & \gate{R_y(\theta_2)}	 & \qw  & \ctrl{0} & \ctrl{0}  & \gate{R_y(\theta_5)} &  \qw
    }
\end{equation*}
\subcaption{}    
\end{minipage}\\
\begin{minipage}[b]{\linewidth}
\centering
\begin{equation*}
     \Qcircuit @C=1.5em @R=1em {
 & \gate{R_y(\theta_0 + 0.36 \pi)}	 & \ctrl{1}  & \qw &  \ctrl{2} &\gate{R_y(\theta_3)} &  \qw \\
  & \gate{R_y(\theta_1 + 0.40 \pi)}	 & \ctrl{0}  & \ctrl{1}& \qw & \gate{R_y(\theta_4)}  &  \qw\\
  & \gate{R_y(\theta_2 + 0.44 \pi)}	 & \qw  & \ctrl{0} & \ctrl{0}  & \gate{R_y(\theta_5)} &  \qw
    }
\end{equation*}
\subcaption{}    
\end{minipage}
\caption{Relationship with Initial Weight. (a) Quantum generators with our proposed distribution as the initial distribution. (b) Quantum generators with our proposed distribution as initial weights.}
\label{fig:init_weight}
\end{figure}
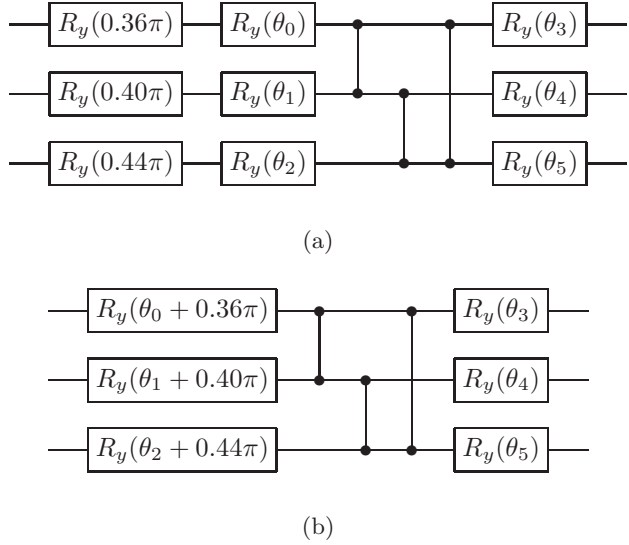

The quantum circuit that generates our initial distributions is realized by $R_y(\theta)$ gates corresponding to the rotation of the Bloch sphere by an angle $\theta$ with the y-axis of the sphere, as shown in Figure \ref{fig:init}.
In addition, the parameter corresponding to the quantum generator weights is the rotation angle $\theta_i$ of the $R_y(\theta_i)$ gate, as shown in Figure \ref{fig:qcircuit}.
Let the rotation angles be $\theta_1$ and $\theta_2$, respectively, then 
\begin{equation}
R_y(\theta_1) \cdot R_y(\theta_2) = R_y(\theta_1 +  \theta_2).
\end{equation}
In other words, $R_y$ gates can be composited with each other, and after compositing, they correspond to a rotational gate $R_y$ with the sum of their respective angles as the rotation angle.
Hence, our initial distribution can be understood as the initial weights of the quantum generators, as shown in Figure \ref{fig:init_weight}.

When the initial distribution is the uniform distribution, the quantum circuit $Init$ consists of three Hadamard gates $H^{\otimes 3}$, as shown in Figure \ref{fig:init_u}, and the operation of a Hadamard gate $H$ on the initial state $\ket{0}$ is
\begin{equation}
H\ket{0} = \frac{1}{\sqrt{2}} \ket{0} + \frac{1}{\sqrt{2}} \ket{1},
\end{equation}
and the operation of $R_y(\frac{\pi}{2})$ on the initial state $\ket{0}$ is
\begin{equation}
R_y(\frac{\pi}{2}) \ket{0} = \frac{1}{\sqrt{2}} \ket{0} + \frac{1}{\sqrt{2}} \ket{1},
\end{equation}
so these two operations are the same for the initial state $\ket{0}$.
Therefore, if $R_y(\frac{\pi}{2})$ is used instead of the Hadamard gate $H$ to create the uniform distribution, this can be composed into the generator parameters as well, corresponding to the fact that the initial weights were chosen as $\frac{\pi}{2}$.
The same interpretation partially holds for the normal distribution since the last column of the circuit in Figure \ref{fig:init_n} is the $R_y$ gate.

In summary, the initial distribution of our method and the uniform distribution can be interpreted as the initial weights of the quantum generator and, simultaneously, as the initial distribution that can be implemented at depth $d=0$.

\section{Conclusion}
We proposed the new method for creating initial distributions by combining single-qubit gates and the classical methods of {\it label replacement} for the qGANs.
Additionally, we proposed new initial distributions that would improve the accuracy of quantum circuit generation using qGANs by using this method and reflecting knowledge of the target probability distributions.
We then tested whether our proposed initial distributions performed better than existing initial distributions by simulation using a classical computer.
%As a result, our method always performed better than uniform distributions consisting of single-qubit gates for all target probability distributions, and it also performed better than more costly normal distributions when targeting log-normal distributions.
The results indicated that our method consistently outperformed uniform distributions composed of single-qubit gates across all target probability distributions. 
Moreover, it surpassed the more costly normal distributions when focusing on log-normal distributions.
In the previous research, a relatively low-cost implementation of the normal distributions was proposed only for the 3-qubit circuit, and we consider that our method is more advantageous for 4-qubit and above.
In addition, deeper quantum circuits may be required to prepare the normal distribution for four or more qubits, potentially eliminating the benefit of circuit generation using qGANs.
In contrast, our initial distribution can consistently be implemented with ``d = 0'' and is robust to the scale of the number of qubits, which is also superior.

We also showed that the proposed and existing initial distributions are composable with the generator weights and correspond to the initial weights.
%As a result, we have shown that our initial and uniform distributions can be implemented at depth $d=0$.
Consequently, we have shown that our initial distributions, as well as uniform distributions, can be implemented at depth $d=0$.

%Our future works are to verify that this method is equally effective for probability distributions other than the log-normal, triangular, and bimodal distributions and to verify its effectiveness when the number of qubits is increased.
Our future works are to examine whether this method maintains its efficacy for probability distributions beyond the log-normal, triangular, and bimodal distributions. We also intend to assess its performance as the number of qubits increases. 
%Furthermore, to achieve the desired level of accuracy, it is imperative to evaluate the impact of the number of measurements and the number of parameters (weights) on accuracy.
In order to achieve the desired level of accuracy,%accuracy, 
it is also necessary to evaluate the impact of the number of measurements and the number of parameters (weights) on accuracy.
%verify the effects of the number of measurements and the number of parameters (weights) on the accuracy.

\section*{Acknowledgment}
YS and RK would like to thank Takayuki Miyadera for the many helpful comments.
YS would like to thank Ikko Hamamura for the technical advice about coding.
This work was supported by JST SPRING Grant Number JPMJSP2110 and JSPS KAKENHI Grant Numbers JP23KJ1178.

\section*{Competing Interests}
All authors declare no financial or non-financial competing interests.

\section*{Data Availability}
The datasets used and analysed during the current study available from the corresponding author on reasonable request.

\section*{Code Availability}
The underlying code for this study is not publicly available but may be made available to qualified researchers on reasonable request from the corresponding author.

\end{document}